\documentstyle[12pt,epic,eepic]{article}
\def\hf{{1\over2}}

\newcommand{\Bbb}{\bf}

\def\gl{{\lower.5ex\hbox{$>$}\atop\raise.5ex\hbox{$<$}}}
\def\lg{{\lower.5ex\hbox{$<$}\atop\raise.5ex\hbox{$>$}}}

\def\half{{1\over2}}
\def\frac#1#2{{#1\over#2}}
\def\ket#1{|#1\rangle}

\def\ds{~\hbox{${\scriptstyle>\atop\raise3pt\hbox{$\scriptstyle<$}}$}~}

\def\infp#1{(#1;q^4)_\infty}

\def\+{\oplus}
\def\gl{\begin{array}{l}\\[-8mm]{\scriptstyle >}\\[-4mm]{\scriptstyle <} \end{array}}

\def\be{\begin{eqnarray}}
\def\ee{\end{eqnarray}}
\def\bea{\begin{eqnarray}}
\def\ena{\end{eqnarray}}
\def\bean{\begin{eqnarray*}}
\def\enan{\end{eqnarray*}}

\def\mref#1{(\ref{#1})}

\def\C{{\bf C}}

\def\F{{\cal F}}

\def\L{\Lambda}

\def\z{\zeta}

\def\bra#1{\langle #1 |}       
\def\ket#1{| #1\rangle}         
\def\br#1{\langle #1 \rangle}   

\def\path{|p\rangle}

\def\End{{\rm End}\,}

\def\id{{\rm id}\,}

\def\com[#1,#2]{\hbox{$[#1,#2]$}}

\def\Phit{\widetilde{\Phi}}  
\def\Psit{\widetilde{\Psi}}  

\def\Phim#1{\mathrel{\mathop{\kern0pt \Phi}\limits^#1}}
\def\phim#1{\mathrel{\mathop{\kern0pt \phi}\limits^#1}}
\def\Psim#1{\mathrel{\mathop{\kern0pt \Psi}\limits^#1}}
\def\Phin#1{\mathrel{\mathop{\kern0pt \Phit}\limits^#1}}
\def\Psin#1{\mathrel{\mathop{\kern0pt \Psit}\limits^#1}}

\def\cd{\cdots}

\newcommand{\bqa}{\begin{eqnarray}}  
\newcommand{\eqa}{\end{eqnarray}}  
\newcommand{\ra}{\rightarrow}  
\newcommand{\lra}{\longrightarrow}
  
\def\L{\Lambda}

\def\half{{1 \over 2}}

\def\bC{{\bf C}}  
\def\D{\Delta}

\def\g{\gamma}  
\def\d{\delta}

\def\ep{\epsilon}

\def\ov{\over}

\def\ed{\end{document}}
\def\ws{\;\;}

\def\ra{\rightarrow}  
  
\def\2pi{1\over 2\pi i}

\def\newline{\hfil\break}

\def\ra{\rightarrow}

\def\sq2{\sqrt{2}}  
\def\sqk2{\sqrt{2(k+2}}  
\def\sqk{\sqrt{k}}

\def\bfig{\begin{figure}}
\def\bfigt{\begin{figure}[top]}
\def\efig{\end{figure}}
\def\bea{\begin{eqnarray}}  
\def\eea{\end{eqnarray}}  
\def\br{\begin{array}}
\def\er{\end{array}}
\def\ea{\end{array}\end{equation}}
\def\bac{\begin{equation}\begin{array}{rll}}
\def\ba{\begin{equation}\begin{array}}
\def\nn{\nonumber} 
 
\def\qp#1#2{({#1}\, ; \, {#2})_{\infty}}
\def\qp4#1{({#1}\, ; \, q^4)_{\infty}}

\def\br#1{({#1};\,x^{2r})_{\infty}}

\def\bR{\bar{R}}

\def\qbinom#1#2{{#1}\atopwithdelims[]{#2}}

\newcommand{\uqp}{U^{\prime}_q (\widehat{sl_2})}

\def\C{{\Bbb C}}

\def\cH{{\cal{H}}}

\def\cF{{\cal{F}}}
\def\cF{{\cal{F}}}

\def\pl{\prod\limits} 

\def\sl{\sum\limits}

\def\ep{\varepsilon}

\def\ss{\subsection}

\def\ot{\otimes}

\def\z{\zeta}
\def\zi{\zeta^{-1}}

\begin{document}
\begin{flushright}
RIMS-1145, DTP-97-17\\
hep-th/9706086 \\[10mm]
\end{flushright}
\begin{center}
{\Large The Monodromy Matrices of the XXZ Model\\[1mm] in 
the Infinite Volume Limit }\\[8mm]
{Tetsuji Miwa\,$^1$ and Robert Weston\,$^2$\\[8mm]
June 1997}
\end{center}
\footnotetext[1]{Research Institute for Mathematical Sciences,
Kyoto University, Kyoto 606, Japan.}
\footnotetext[2]{Department of Mathematical Sciences,
University of Durham, Science Labs,\\[-1mm]
\hspace*{7mm}South Rd, Durham DH1 3LE, U.K. \ \ Email: R.A.Weston@durham.ac.uk}

\begin{abstract}
\noindent 
We consider the XXZ model in the infinite volume limit 
with spin $\half$ quantum space and higher
spin auxiliary space. Using perturbation theory arguments, we 
relate the half infinite transfer matrices of this class of models
to certain $U_q(\widehat{sl_2})$ intertwiners introduced by
Nakayashiki. We construct the monodromy matrices, and show
that the one with spin $1$ auxiliary space
gives rise to the $L$ operator.
\end{abstract}
\setcounter{equation}{0}
\setcounter{section}{0}
\section{Introduction}
In this paper, we revisit the $U_q(\widehat{sl}_2)$ symmetry
of the XXZ Hamiltonian in the infinite volume limit. We show
that the monodromy matrix, in the sense used in the quantum inverse scattering
method, gives rise to the $L$ operator representing the
level $0$ action
of $U'_q(\widehat{sl}_2)$ on the physical space of states.
In this construction, we take the auxiliary space
for the monodromy matrix to be the spin $1$ $U'_q(\widehat{sl}_2)$ module
$V^{(2)}_\zeta$, i.e., the $3$-dimensional evaluation module with
spectral parameter $\zeta$. The monodromy matrix acts on the quantum
space ${\cal F}$, which is the $N\rightarrow\infty$ limit
of the $N$-fold tensor product of spin $\half$ $U'_q(\widehat{sl}_2)$
modules $V^{(1)}$.

If we take the spin $\half$ module $V^{(1)}_\zeta$
as the auxiliary space, and consider
the trace of the monodromy matrix acting on a finite, say
$N$-fold, tensor product,
we obtain the transfer matrix $T^{(1)}_N(\zeta)$ for the six vertex model
of size $N$.
$T^{(1)}_N(\zeta)$ form a commuting family of operators which contains
the XXZ Hamiltonian.
In general, for any positive integer $m$, one can
define a family of operators $T^{(m)}_N(\zeta)$ which commute with 
$T^{(1)}_N(\zeta)$ and among themselves, by choosing the spin ${m\over2}$
auxiliary space $V^{(m)}_\zeta$. However,
these are not new operators because the fusion relation expresses
them as polynomials in $T^{(1)}_N(\zeta)$ with suitably shifted parameters.
On the other hand, the components of the monodromy matrix
other than its trace do not commute among themselves,
and in fact if $m=1$, they obey the commutation relation of the $L$ operator.

The situation is different in the infinite  volume limit.
This is because one must choose appropriate boundary conditions
in order to have a well-defined limit of the monodromy matrix.
For example, suppose we take the spin $\half$ auxiliary space.
We restrict our discussion to the massive case, i.e., $-1<q<0$.
The dominant Boltzmann weight in this regime is the $c$ weight
(in the usual terminology of the six vertex model). Let us normalize
it to be $1$. The weights $a$ and $b$ are then small.
To have a non-zero contribution, we must take all but a finite number
of constituents
of the monodromy matrix to be the $c$ weight.
Therefore, the choice of the boundary condition in the quantum space uniquely
fixes the choice in the auxiliary space.
In other words, the total spin
of the auxiliary space at the boundaries effectively changes to $0$.

We denote the monodromy matrix in this sense,
acting on the infinite tensor product, by $T^{(1)}(\zeta)$,
and call it the monodromy matrix in the infinite volume limit.
The XXZ Hamiltonian is obtained from the first derivative
of $T^{(1)}(\zeta)$ at $\zeta=1$. In short, the
transfer matrix in the infinite volume limit
is nothing but the monodromy matrix with the spin $\half$
auxiliary space.

In general, 
the effective total spin of
the monodromy matrix at the boundaries is equal to $(n-1)/2$ 
 if we take
the auxiliary space to be $V^{(n)}_\zeta$.
We denote this operator by
$T^{(n)}(\zeta)=\Bigl(T^{(n)}_{l,l'}(\zeta)\Bigr)_{l,l'=0,\ldots,n-1}$.
The main result of this paper is to show that 
the monodromy matrix $T^{(2)}(\z)$ can be interpreted as the $L$ operator.

The physical space of states for the XXZ model consists of
the vacuum vectors and the multi-particle states. Particles have spin $\half$.
Namely, they transform according to the $2$-dimensional evaluation
representation of $U'_q(\widehat{sl}_2)$ and constitute a space of states
which is isomorphic to the tensor product of $V^{(1)}_{\xi_i}$
$(i=1,\ldots,m)$. This is called the particle picture of the space of states.
The action of the transfer matrix $T^{(1)}(\zeta)$ is diagonalized in
the particle picture. It is $2^m$-fold degenerate on each $m$-particle space
$V^{(1)}_{\xi_1}\otimes\cdots\otimes V^{(1)}_{\xi_m}$
with a given set of spectral parameters $\{\xi_1,\ldots,\xi_m\}$.
In order to resolve this degeneracy, we need the $U'_q(\widehat{sl}_2)$
symmetry. We calculate the action of the monodromy matrix
$T^{(n)}(\zeta)$ on $V^{(1)}_{\xi_1}\otimes\cdots\otimes V^{(1)}_{\xi_m}$
explicitly. The result is
essentially equal to the action of the monodromy matrix of size $m$
with spin ${(n-1)/ 2}$ auxiliary space.

Now, let us come to the representation theoretical content of the story.
The two key observations in the series of works 
[1-6] on the XXZ model are that the half
infinite tensor
product of $V^{(1)}$ can be identified with the level $1$ integrable
highest weight representations ${\cal H}=V(\Lambda_0)\oplus V(\Lambda_1)$
of $U_q(\widehat{sl}_2)$, and that the half infinite transfer matrix acting on it
is identified with an intertwiner called the type I vertex operator.
The space of states is given as ${\cal F}={\cal H}\otimes{\cal H}^*$.
This is called the local picture of the space of states.

In \cite{Nak96}, a new class of intertwiners is introduced, which
generalises the type I vertex operator. For each integer $n\geq0$,
we consider Nakayashiki's intertwiner
\bea
\Phi^{(n)}(\zeta):V^{(n)}_\zeta\otimes V(\Lambda_i)\rightarrow
V(\Lambda_{1-i})\otimes V^{(n+1)}_\zeta.
\nn\ena
We show (up to a few orders in $q$) that
the infinite volume limit of the half transfer matrix with
auxiliary space $V^{(n+1)}_\zeta$ is identified with $\Phi^{(n)}(\zeta)$
for $n=0,1$, and conjecture that this statement is valid to all orders for
all $n$. 
From this follows the representation theoretical definition of the
monodromy matrix $T^{(n+1)}(\z)$ (see \mref{Tdef}).
We then compute the action of $T^{(n)}(\z)$
on the space of states in the particle picture, and thereby
derive the commutation relations of $T^{(n)}(\z)$. Finally, we derive
the fusion relation which expresses $T^{(n)}(\z)$ in terms of
$T^{(2)}(\z)$ and $T^{(1)}(\z)$ with suitably shifted parameters.
\setcounter{equation}{0}
\setcounter{section}{1}
\section{Half Transfer Matrices}\label{sechtm}
Consider the six vertex model specified by the following Boltzmann weights:
\bea\label{abc}
&&{\tilde a}={\tilde R}^{0,0}_{0,0}={\tilde R}^{1,1}_{1,1}={1-q^2\z^2\over\z(1-q^2)},\nonumber\\
&&{\tilde b}={\tilde R}^{0,1}_{0,1}={\tilde R}^{1,0}_{1,0}={q(1-\z^2)\over\z(1-q^2)},\nn\\
&&{\tilde c}={\tilde R}^{0,1}_{1,0}={\tilde R}^{1,0}_{0,1}=1.\nonumber
\ena
We restrict our consideration to the  parameter region 
$-1<q<0$ and $1<\z<-q^{-1}$. The $R$-matrix
${\tilde R}=\Bigl({\tilde R}^{k'_1,k'_2}_{k_1,k_2}\Bigr)$ acts on
the tensor product $\C^2\otimes \C^2$. Following
the terminology of the quantum inverse scattering method,
we call the first component of the tensor product
the auxiliary space, and the second component the quantum space.
The monodromy matrix $T_N{}_{l'}^l$ ($l,l'=0,1$)
is an operator acting on
the $N$-fold tensor product of the quantum space
$\C^2=\C u^{(1)}_0\oplus\C u^{(1)}_1$:
\nn\bea
T_N{}_{l'}^l(u^{(1)}_{k_N}\otimes\cdots\otimes u^{(1)}_{k_1})
&=&\sum_{k'_1,\ldots,k'_N\atop l_1,\ldots,l_{N-1}}
{\tilde R}^{l_1,k_1}_{l',k'_1}\cdots
{\tilde R}^{l,k_N}_{l_{N-1},k'_N}
u^{(1)}_{k'_N}\otimes\cdots\otimes u^{(1)}_{k'_1}.
\nn\ena

In this section, we use a small $q$ expansion to compare the $N\ra
\infty$ limit of
the monodromy matrix of size $N$ with the level $1$ intertwiners.
We take the limit keeping the spin variable at one end fixed and
changing the one at the other end according to the choice of boundary condition.
This limit gives the action of the half transfer matrix in the infinite
volume limit.
\subsection{The half infinite tensor product}
We make the identification of the half infinite tensor product
$\lim_{N\rightarrow\infty}(\C^2)^{\otimes N}$
with the level $1$ $U_q(\widehat{sl}_2)$ module
${\cal H}=V(\Lambda_0)\oplus V(\Lambda_1)$ in the way discussed in
\cite{FM,DFJMN,JM}.
For $i=0,1$, consider the set of paths ${\cal P}^{(i)}$ consisting of
sequences of $0,1$, denoted by $|p\rangle_{(i)}=\{p(j)\}_{j\ge1}$,
which satisfy
the boundary condition $p(j)=(1-(-1)^{i+j})/2$ for sufficiently large $j$.
Consider the vector space ${\cal H}^{(i)}$ spanned by the formal expressions
$\sum_{p\in{\cal P}^{(i)}}c(p)|p\rangle_{(i)}$, where $c(p)$ is a formal power
series in $q$. The statement (though not a mathematical theorem)
is that there is an embedding $\kappa$ of the irreducible
highest weight module $V(\Lambda_i)$ into ${\cal H}^{(i)}$ such that
the action of $U_q(\widehat{sl}_2)$ on $V(\Lambda_i)$
is induced from the formal action of $U'_q(\widehat{sl}_2)$
on the half infinite tensor product given by the coproduct (\ref{coprod}).

For example, the path expansion of the highest weight vector
$|\Lambda_0\rangle$ reads
\bea\label{PATH}
\kappa(|\Lambda_0\rangle)
&=&\cd-q\sum\cd(2)\cd\nonumber\\
&&+q^2\biggl(\sum\cd(2)\cd(2)\cd
+2\sum\cd(4)\cd\biggr)\nonumber\\
&&+q^3\biggl(\sum\cd(2)+2\sum_{k\ge1}\cd(2)(k)-\sum\cd(1)(2)(1)\cd\biggr)
+O(q^4).\nonumber\\
\ena
The notation is as follows: We call an alternating
sequence of $0,1$ of maximal length a domain. We decompose a path into domains.
In the above formula, we denote by $(k)$ a domain of length $k$.
The symbol $\cd$, when it is placed at the leftmost end, means
an infinite domain. Otherwise $\cd$ means a domain of undetermined
length. The length of such a domain can be any strictly positive integer.
In addition, the length can be $0$ if $\cd$ is placed at the rightmost end.

The path expansion of the vectors in $V(\Lambda_1)$ are obtained by
the $0\leftrightarrow1$ symmetry. In particular, the expansion
of $\kappa(\ket{\L_1})$ in the above notation is exactly the same
as (\ref{PATH}).
\subsection{The perturbative action of the half transfer matrix}
For each $N$, we set
\bea
{\cal P}^{(i)}_N&=&\{|p\rangle_{(i)};p(j)=(1-(-1)^{i+j})/2\hbox{ if $j>N$}\}.
\nn\ena
We denote by $\rho_N$ the projection of ${\cal H}^{(i)}$ to
the vector space spanned by ${\cal P}^{(i)}_N$.
We define an operator ${\tilde \Phi}_{N,k}(\zeta)$ acting from
$\rho_N({\cal H}^{(i)})$ to $\rho_N({\cal H}^{(1-i)})$ by
\bea\label{HTM}
{\tilde \Phi}_{N,k}(\zeta)|p\rangle_{(i)}
&=&\sum_{|p'\rangle_{(1-i)}\in{\cal P}^{(1-i)}_N}
\sum_{k_1,\ldots,k_{N-1}}
\prod_{1\leq j\leq N}{\tilde R}^{k_j,p(j)}
_{k_{j-1},p'(j)}(\z,q)|p'\rangle_{(1-i)}
\ena
where
\bea
k=0,1,\quad k_0=k, \quad
k_N=\half(1-(-1)^{i+N+1}).
\nn\ena
Since $\tilde c=1$, the matrix element of
the half transfer matrix between $|p\rangle_{(i)}$
and $|p'\rangle_{(1-i)}$ 
is uniquely determined by the formula (\ref{HTM})
by taking a sufficiently large $N$. If we take
$|p'\rangle_{(i)}$ instead of $|p'\rangle_{(1-i)}$ in (\ref{HTM}),
the matrix element
vanishes in the limit $N\rightarrow\infty$ because $|a|,|b|<1$.
Note also that for given $|p\rangle_{(i)}\in{\cal P}^{(i)}$ and
$|p'\rangle\in{\cal P}^{(1-i)}$, the $k_j$ $(1\leq j\leq N-1)$
are uniquely determined. Only one term in the second sum of (\ref{HTM})
is non-vanishing.

Now, we apply formula (\ref{HTM}) to a vector which belongs to
${\cal H}^{(i)}$. It is an infinite linear combination of $|p\rangle$, and
the coefficient of $|p'\rangle$ in the right-hand side
must be summed up with respect to these $|p\rangle$.
This sum diverges in the limit $N\rightarrow\infty$.
To have a finite sum we need to renormalize the Boltzmann weights.
We use the normalization \mref{RMAT} of the $R$-matrix such that
the partition function is $1$ (see \cite{JM}).
The expansion of the weights $a$, $b$, $c$
reads as
\bea\label{ABC}
a&=&\z^{-1}+q^2(\z^{-3}-\z)+O(q^4),\nonumber\\
b&=&q(\z^{-1}-\z)+q^3(\z^{-3}-\z^{-1})+O(q^4),\\
c&=&1+q^2(\z^{-2}-1)+O(q^4).\nonumber
\ena
We define $\Phi_{N,k}(\zeta)$ as in (\ref{HTM}) but with
${\tilde a}$, ${\tilde b}$, ${\tilde c}$ replaced by $a$, $b$, $c$.

Let $\Phi_k(\z)$ be the type I vertex operator (see
(\ref{EVO}) and \mref{TYPI}).
We conjecture
\bea
\hbox{\rm lim}_{N\rightarrow\infty}
{1\over \mu(\z,q)}
\Phi_{N,k}(\zeta)\circ\rho_N\circ\kappa
&=&\kappa\circ\Phi_k(\z)
\label{FINI}
\ena
where $\mu(\z,q)$ is a series in $q$ with Laurent polynomials in $\z$ as
coefficients.

Let us check (\ref{FINI}) on the vector $|\Lambda_0\rangle$:
\bea
\hbox{\rm lim}_{N\rightarrow\infty}
{1\over \mu(\z,q)}
\Phi_{N,k}(\z)\circ\rho_N\circ\kappa(|\Lambda_0\rangle)
&=&\kappa\circ\Phi_k(\z)(|\Lambda_0\rangle).
\label{CHK}
\ena
We find
\bea
\mu(\z,q)&=&1+q^2(\z^{-2}-1)+O(q^4).
\nn\ena
First, we will check the coefficients of $\cd$, $\cd(2)$ and
$\cd(2)(m)$ $(m\ge1)$ in (\ref{CHK}) with $k=1$.
The coefficients of these terms in
$\Phi_{N,1}(\zeta)\circ\rho_N\circ\kappa(|\Lambda_0\rangle)$ are, modulo $O(q^4)$,
\bea
&&c^N-(N-1)qabc^{N-2},\nn\\
&&(-q+2q^3)a^2c^{N-2}+abc^{N-2}+(N-3)q^2a^3bc^{N-4}-(N-3)qa^2b^2c^{N-4},\nn\\
&&(-q+2q^3)a^2c^{N-2}+(1+2q^2)abc^{N-2}-qb^2+(N-4)q^2a^3bc^{N-4}
-(N-4)qa^2b^2c^{N-4},\nonumber
\ena
respectively. Using (\ref{ABC}) we obtain
\bea\label{PERT}
&&\hbox{\rm lim}_{N\rightarrow\infty}
\Phi_{N,1}(\zeta)\circ\rho_N\circ\kappa(|\Lambda_0\rangle)=
\biggl(1+q^2(\z^{-2}-1)+O(q^4)\biggr)\times\nonumber\\
&&\biggl(\cd+(-q+q^3)\cd(2)+(-q+2q^3)\sum_{m\ge1}\cd(2)(m)+\cdots
\nonumber\\
&&+\z^2q^3\cd(2)+\cdots\biggr).
\ena

Note that
\bea\label{ACTI}
\Phi(\z)|\Lambda_0\rangle
&=&|\Lambda_1\rangle\otimes u^{(1)}_1
-q\z f_1|\Lambda_1\rangle\otimes u^{(1)}_0
+{q^4\z^2\over1+q^2}f_0f_1|\Lambda_1\rangle\otimes u^{(1)}_1+\cdots.
\ena
Following the method of \cite{FM}, we obtain
\bea
\kappa(f_0f_1|\Lambda_1\rangle)
&=&(q^{-1}+\cdots)\cd(2)+\cdots.
\nn\ena
Therefore, the result (\ref{PERT}) is consistent with the conjecture.

Next we check the case $k=0$ in (\ref{FINI})
for the coefficient of $\cd(1)$.
It is, modulo $O(q^4)$,
\bea
&&bc^{N-1}+(-q+q^3)ac^{N-1}+(N-2)q^2a^2bc^{N-3}-(N-2)qab^2c^{N-3}.
\nn\ena
Using (\ref{ABC}) we obtain
\bea
\hspace*{-4mm}\hbox{\rm lim}_{N\rightarrow\infty}
\Phi_{N,0}(\zeta)\circ\rho_N\circ\kappa(|\Lambda_0\rangle)=
\biggl(1+q^2(\z^{-2}-1)+O(q^4)\biggr)
\biggl(-q(1-q^2)\z\cd(1)+\cdots\biggr).\label{AST}
\ena
On the other hand, we have
\bea
\kappa(f_1|\Lambda_1\rangle)
&=&(1-q^2)\cd(1)+\cdots,
\nn\ena
in (\ref{ACTI}). Therefore, the result (\ref{AST}) is
consistent with the conjecture (\ref{FINI}).
\subsection{The case with spin 1 auxiliary space}
Let us calculate similar quantities for the spin $1$ auxiliary space.
We take the weight vectors $u^{(2)}_j$ $(j=0,1,2)$ of the three dimensional 
$U'_q(\widehat{sl}_2)$ module $V^{(2)}_\z$ as in (\ref{BASE}).
This is the choice such that
the matrix elements of the spin $(1,\half)$ $R$-matrix are
\bea
&&{\tilde A}={\tilde R}^{0,0}_{0,0}={\tilde R}^{2,1}_{2,1}={1\over\sqrt{1+q^2}}
{1-q^3\z^2\over\z(1-q^2)},\nn\\[1mm]
&&{\tilde B}={\tilde R}^{0,1}_{0,1}={\tilde R}^{2,0}_{2,0}={1\over\sqrt{1+q^2}}
{q(q-\z^2)\over\z(1-q^2)},\nn\\[1mm]
&&{\tilde C}={\tilde R}^{1,0}_{1,0}={\tilde R}^{1,1}_{1,1}={1\over\sqrt{1+q^2}}
{q(1-q\z^2)\over\z(1-q^2)},\nn\\[2mm]
&&{\tilde D}={\tilde R}^{2,0}_{1,1}={\tilde R}^{0,1}_{1,0}
={\tilde R}^{1,0}_{0,1}={\tilde R}^{1,1}_{2,0}=1.\nn
\ena
We set
\bea
{\tilde \Phi}^{(1)}_{N,l,k}(\zeta)|p\rangle_{(i)}&=&
\sum_{|p'\rangle_{(1-i)}\in{\cal P}^{(1-i)}_N}\sum_{k_1,\ldots,k_{N-1}}
\prod_{1\leq j\leq N}{\tilde R}^{k_j,p(j)}_{k_{j-1},p'(j)}(\z,q)|p'\rangle
_{(1-i)},
\nn\ena
where
\bea
l=0,1,\quad k=0,1,2, \quad k_0=k, \quad
k_N=l+\half(1-(-1)^{i+N+1}).
\nn\ena

Note that the ${\tilde D}$ weight is dominating others. 
Therefore, if we choose $|p\rangle_{(i)}$ in ${\cal P}^{(i)}$, we must choose
$|p'\rangle_{(1-i)}$ in ${\cal P}^{(1-i)}$. This is the same as in the spin $\half$
case. A new phenomenon is that there are two choices of the value of $k_N$
corresponding to $l=0,1$. In order to obtain finite results
we define $\Phi^{(1)}_{N,l,k}(\zeta)$ by using the normalization
(\ref{RMAT})
for the $R$-matrix. The overall normalization factor (corresponding to the
$\mu(\z,q)$
in the case of the spin $\half$ auxiliary space)  now depends upon
whether $N$ is odd or even.
The conjecture is
\bea
&&\hbox{\rm lim}_{N\rightarrow\infty}
{1\over \nu_N(\z,q)}
\Phi^{(1)}_{N,l,k}(\zeta)\circ\rho_N\circ\kappa
=\kappa\circ\Phi^{(1)}_{l,k}(\z),\nn\\
&&\nu_N(\z,q)=1-\biggl({N\over2}-\left[{N\over2}\right]\biggr)q^2
+q^3\z^{-2}+O(q^4),\nn
\ena
where we use Nakayashiki's operator $\Phi^{(1)}(\z)$ (see (\ref{gint}))
in the right-hand side. In the above formula
the symbol $\left[{N\over2}\right]$ means the integer part of $N/2$.

The following are the supporting calculations.
Let us check the case $l=1,k=2$ on the vector $|\Lambda_0\rangle$.
We have
\bea
\Phi^{(1)}_{1,2}(\z)|\Lambda_0\rangle&=&
|\Lambda_1\rangle+\cdots.
\nn\ena
The coefficients of $\cd$, $\cd(2)$, $\cd(2)(m)$ $(m\ge2,\hbox{ even})$
and $\cd(2)(m)$ $(m\ge1,\hbox{ odd})$ in
$\Phi^{(1)}_{N,1,2}(\zeta)\circ\rho_N\circ\kappa(|\Lambda_0\rangle)$ are, modulo $O(q^4)$,
\bea
D^N-\left[{N\over2}\right]qABD^{N-2}
-\left[{N-1\over2}\right]qC^2D^{N-2}
&\equiv&\nu_N(\z,q),\nonumber\\[2mm]
ABD^{N-2}-qACD^{N-2}-\biggl(\left[{N\over2}\right]-1\biggr)qA^2B^2D^{N-4}
&\equiv&(-q+q^3)\nu_N(\z,q),\nonumber\\[2mm]
-qBCD^{N-2}
+ABD^{N-2}-qACD^{N-2}-\biggl(\left[{N\over2}\right]-1\biggr)qA^2B^2D^{N-4}
&\equiv&(-q+2q^3)\nu_N(\z,q),\nonumber\\[2mm]
C^2D^{N-2}-qBCD^{N-2}+(-q+2q^3)D^N
-qACD^{N-2}+\left[{N\over2}\right]q^2ABD^{N-2}
&\equiv&(-q+2q^3)\nu_N(\z,q),\nonumber
\ena
respectively. These are consistent with the conjecture.

Let us check the case $l=0,k=1$ on the vector $|\Lambda_0\rangle$.
We have
\bea
\Phi^{(1)}_{0,1}(\z)|\Lambda_0\rangle
=(1-{q^2\over2})|\Lambda_1\rangle+\cdots.
\nn\ena
The perturbative calculation gives
the coefficient of $\cd$ in
$\Phi^{(1)}_{N,0,1}(\zeta)\circ\rho_N\circ\kappa(|\Lambda_0\rangle)$ to be, modulo $O(q^4)$,
\bea
D^N-\left[{N\over2}\right]qC^2D^{N-2}-\left[{N-1\over2}\right]qABD^{N-2}
&\equiv&(1-{q^2\over2})\nu_N(\z,q).
\nn\ena
This is consistent with the conjecture.

Based on the perturbative checks in this and the previous subsection, we
conjecture that the half transfer matrix with spin $(n+1)/2$
auxiliary space is given by Nakayashiki's operator $\Phi^{(n)}(\zeta)$.
More specifically, we define the half transfer matrix with spin
$(n+1)/2$ by
\be \Phi_{N\,l,k}^{(n)}(\z)\ket{p}_{(i)}=
\sum_{|p'\rangle_{(1-i)}\in{\cal P}^{(1-i)}_N}
\sl_{k_1,\cdots,k_{N-1}}
\pl_{N\leq j\leq 1} R^{(n+1,1)}
(\z)^{k_j,p(j)}_{k_{j-1},p'(j)}\ket{p'}_{(1-i)},\label{def1}\ee
where $0\leq l\leq n$, $0\leq k\leq n+1$,
$k_0=k$, $k_N=l+\hf(1-(-1)^{N+i+1})$,
and where $R^{(n+1,1)}(\z)$ is given in (\ref{RMAT}).
Our conjecture is that
$ \lim_{N\ra \infty} 
\Phi_{N\, l,k}^{(n)}(\z) \circ \rho_N\circ \kappa$ is proportional to
$\kappa\circ \Phi_{l,k}^{(n)}(\z)$, where 
$\Phi_{l,k}^{(n)}(\z)$ is given by (\ref{gint}) and \mref{gint2}. 
\subsection{The monodromy matrix in the infinite volume limit}
Consider the monodromy matrix of size $2N$,
\bea
T_{2N\, l,l'}^{(n+1)}(\z)^{p(N),\cdots,p({1-N})}_{p'(N),\cdots,p'({1-N})}=
\sl_{k_{N-1},\cdots,k_{1-N}}
\pl_{N\leq j\leq 1-N} R^{(n+1,1)}(\z)^{k_j,p(j)}_{k_{j-1},p'(j)},\nn\ena
where 
$k_N=l+\hf(1-(-1)^{N+i+1})$, $k_{-N}=l'+\hf(1-(-1)^{N+i+1})$,
$0\leq l,l'\leq n$ and $p(j),p'(j)=0,1$.
Using \mref{def1}, and the symmetry 
$R^{(n+1,1)}(\z)^{a,b}_{c,d}=R^{(n+1,1)}(\z)^{n+1-c,1-d}_{n+1-a,1-b}$,
we can rewrite this as
\be
T_{2N\, l,l'}^{(n+1)}(\z)^{p(N),\cdots,p({1-N})}_{p'(N),\cdots,p'({1-N})}=
\sl_{k=0}^{n+1}
\Phi_{N\, l,k}^{(n)}(\z)^{p(1),\cdots,p(N)}_{p'(1),\cdots,p'(N)} \otimes 
\Phi_{N\, n-l',n+1-k}^{(n)}
(\z)^{1-p'(0),\cdots,1-p'({1-N})}_{1-p(0),\cdots,1-p({1-N})} 
.\nn\ee

This observation, together with the conjectural form of $ \lim_{N\ra
\infty}\Phi_{N\, l,k}^{(n)}(\z)$,
motivates us to define the monodromy matrix in the infinite volume limit,
$T_{l,l'}^{(n+1)}(\z)\in
\End(\cH\ot \cH^*)$, as 
\be T_{l,l'}^{(n+1)}(\z) = g^{(n)} \sl_{k=0}^{n+1} 
\Phi^{(n)}_{l,k} (\z) \ot \Phi^{(n)\ t}_{n-l',n+1-k}(\z),\label{Tdef}\ee
where $g^{(n)}$ is a constant which appears in
the next section. This is a generalisation of (7.8)
(the case $n=0$) in \cite{JM}.
\setcounter{equation}{0}
\setcounter{section}{2}
\section{Level-one Intertwiners}\label{secint}
In this section, we discuss level-one intertwiners of the algebra
$U'=\uqp$ generated by $e_i,f_i,t_i$ (i=0,1).
Unless otherwise stated,
all notational conventions are those of \cite{JM}. We choose the
coproduct $\D$ to be
\be \D(t_i)=t_i \ot t_i, \quad \D(e_i)=e_i\ot 1+ t_i\ot
e_i,\quad \D(f_i)=f_i\ot t_i^{-1}+ 1\ot f_i.
\label{coprod}\ee
We need two types of $U'$ modules for our analysis; level-one
highest weight modules, and level-zero evaluation modules.
Level-one highest weight modules $V(\L_j)$ $(j=0,1)$ are generated
by the highest weight vector $v_{\L_j}$ which
obeys $e_i v_{\L_j}=0$, $t_i v_{\L_j}=q^{\d_{i,j}}
v_{\L_j}$ and $f_i^{\d_{i,j}+1} v_{\L_j}=0$.
\ss{Evaluation modules}
We use a principally specialised evaluation module 
$V^{(n)}_{\z}$, with weight vectors $u^{(n)}_l$.
The action of $U'$ on
$V^{(n)}_{\z}$ is given by
 \bac
&& f_1 u_l^{(n)} = \z^{-1} b^{(n)}_l u_{l+1}^{(n)}, \quad
e_1 u_l^{(n)} = \z b^{(n)}_{n-l} u_{l-1}^{(n)},\quad t_1 u_l^{(n)}=q^{n-2l}
u_l^{(n)},\label{BASE}\\
&& f_0  u_l^{(n)}= \z^{-1}  b^{(n)}_{n-l} u_{l-1}^{(n)},
\quad
 e_0 u_l^{(n)}= \z b^{(n)}_l u_{l+1}^{(n)},\quad
t_0 =t_1^{-1},
\nn\ea
where $b_{l}^{(n)}= q^{(-n+2l+1)/2} ([l+1][n-l])^{1/2}$. 
In this basis we have an isomorphism
\be
V_{-q^{-1}\zeta}\rightarrow V^{*a}_\zeta,
\quad u^{(n)}_l\mapsto u^{(n)*}_{n-l},
\nn\ee
where $\langle u^{(n)*}_l,u^{(n)}_k\rangle=\delta_{lk}$.

We now consider the structure of tensor products of such modules.
First we note that $V^{(1)}_{q\z} \ot V_{\z}^{(1)}$ has a one-dimensional
$U'$ submodule
\be M \hookrightarrow  V^{(1)}_{q \z} \ot V_{\z}^{(1)}, \quad{\rm where}
\ws M=\bC(u_0^{(1)} \ot u_1^{(1)} - u_1^{(1)} \ot u_0^{(1)}).\nn\ee
Letting 
$N_j= V^{(1)}_{q^{n-1 \ov 2}\z} \ot V^{(1)}_{q^{n-3 \ov 2}\z}
\ot \cdots \ot M \ot \cdots \ot
V^{(1)}_{q^{1-n \ov 2}\z} $,
where $M$ is
the corresponding sub-module in the $(j,j+1)$ position, we have
\be V_{\z}^{(n)} \simeq  V_{q^{n-1\ov 2} \z}^{(1)} \ot 
V_{q^{n-3\ov 2} \z}^{(1)} \ot \cdots \ot
V_{q^{{1-n\ov 2}} \z}^{(1)} /\ \sl_{j=1}^{n-1} N_j.\nn\ee
The $U'$ linear projection $\pi^{(n)}:V_{q^{n-1\ov 2} \z}^{(1)} \ot
V_{q^{n-3\ov 2} \z}^{(1)} \ot \cdots \ot
V_{q^{{1-n\ov 2}} \z}^{(1)} \lra V_{\z}^{(n)}$ is given by
\be
\pi^{(n)} (u^{(1)}_{l_1}\ot u^{(1)}_{l_2}\ot\cdots \ot u^{(1)}_{l_n})=
\g^{(n)}_{l_1+\cdots +l_n} u^{(n)}_{l_1+l_2+\cdots +l_n} ,\nn\ee
where $\g^{(n)}_l= {\qbinom{n}{l}}^{-\hf}$.

On the other hand, $V_{q^{1-n\ov 2} \z}^{(1)} \ot
 V_{q^{3-n\ov 2}\z}^{(1)}\ot\cdots \ot
V_{q^{{n-1\ov 2}} \z}^{(1)}$ has a $U'$ sub-module $V^{(n)}_{\z}$.
The $U'$ linear embedding $\iota^{(n)}:V_{\z}^{(n)} 
\hookrightarrow V_{q^{1-n\ov 2} \z}^{(1)} \ot V_{q^{3-n\ov 2} \z}^{(1)} \ot \cdots \ot
V_{q^{{n-1\ov 2}} \z}^{(1)}$ is given by
\be
\iota^{(n)}( u_l^{(n)}) = \gamma^{(n)}_l\tilde{u}^{(n)}_l,\quad
{\rm where}\ws
\tilde{u}^{(n)}_l=\sl_{l_1+\cdots+l_n=l} 
u^{(1)}_{l_1}\ot u^{(1)}_{l_2}\ot\cdots\ot u^{(1)}_{l_n}.\nn\ee

We construct certain $R$-matrices associated with our evaluation
modules. The $R$-matrix $\bR^{(n,m)}(\z_1/\z_2) : 
V_{\z_1}^{(n)} \ot V_{\z_2}^{(m)} \lra V_{\z_1}^{(n)}
\ot V_{\z_2}^{(m)}$ is defined up to a normalisation by the
requirement that $\bR^{(n,m)}(\z_1/\z_2) \D(x) = \D'(x)
\bR^{(n,m)}(\z_1/\z_2)$ on $V_{\z_1}^{(n)} \ot V_{\z_2}^{(m)}$.
Here, $x\in U'$ and $\D'(x)$ is defined in \cite{JM}.
On $V^{(n)}_{\z_1}\ot V^{(m)}_{\z_2} \ot V^{(p)}_{\z_3}$, we have
the Yang-Baxter equation
\be \bR^{(m,p)}(\z_2/\z_3) \bR^{(n,p)}(\z_1/\z_3) \bR^{(n,m)}(\z_1/\z_2) 
=  \bR^{(n,m)}(\z_1/\z_2) \bR^{(n,p)}(\z_1/\z_3)
\bR^{(m,p)}(\z_2/\z_3).
\label{YB}\ee
We define components by
\be
\bR^{(n,m)} (\z_1/\z_2) u^{(n)}_l \ot u^{(m)}_{k} =
\sl_{l',k'} u^{(n)}_{l'} \ot u^{(m)}_{k'} 
\bR^{(n,m)}(\z_1/\z_2)^{l,k}_{l',k'},\nn\ee
and fix the normalisation by requiring that
$\bR^{(n,m)}(\z)^{0,0}_{0,0}=1$.
The simplest way to obtain $\bR^{(n,1)}(\z)$ is through
the fusion technique. This gives
\be
\bR^{(n,1)}(\z)^{l,k}_{l',k'}={\g^{(n)}_{l'} \ov \g^{(n)}_l}\hspace*{0mm}
\sl_{l'_1+\cdots+l'_n=l',k_1,\cdots,
k_{n-1}} 
\hspace*{-10mm}
\bR(\z q^{n-1\ov 2})^{l_1,k_1}_{l'_1,k'} 
\bR(\z q^{n-3\ov 2})^{l_2,k_2}_{l'_2,k_1} \,
\cdots \bR(\z q^{1-n\ov 2})^{l_n,k}_{l'_n,k_{n-1}},
\label{fusion}\ee
where $l_1+\cdots+l_n=l$, and $\bR(\z)=\bR^{(1,1)}(\z)$.
Explicitly, we have
\be \bR^{(n,1)}(\z)^{l,0}_{l,0}= 
{q^l(1-q^{n+1-2l}\z^2)\ov 1-q^{n+1}\z^2},\quad
\bR^{(n,1)}(\z)^{l,1}_{l+1,0}= ([n-l][l+1])^{\hf} q^{n-1\ov 2}
{(1-q^2) \z \ov 1-q^{n+1} \z^2}.\nn\ee
Other components are given by the symmetries
$\bR^{(n,1)}(\z)^{a,b}_{c,d}=\bR^{(n,1)}(\z)^{n-a,1-b}_{n-c,1-d}=
\bR^{(n,1)}(\z)_{a,b}^{c,d}$. The other $R$-matrix to which we shall
refer later on is $\bR^{(1,n)}(\z)$. This is given by
$\bR^{(1,n)}(\z)^{a,b}_{c,d}=\bR^{(n,1)}(\z)^{b,a}_{d,c}$.
Finally, we shall use the normalised R-matrices 
\bac R^{(n,1)}(\z)&=&\bR^{(n,1)}(\z)/\kappa^{(n)}(\z),\quad
R^{(1,n)}(\z)=\bR^{(1,n)}(\z)/\kappa^{(n)}(\z), \quad {\rm where}\ws \\[2mm]
\kappa^{(n)}(\z)&=&\z {\qp4{q^{3+n}\z^2} \qp4{q^{1+n} \z^{-2}} \ov
\qp4{q^{3+n}\z^{-2}} \qp4{q^{1+n} \z^{2}} }.\label{RMAT}\ea
With this factor, $R^{(n,1)}(\z)$ 
enjoys the properties of unitarity
$\sl_{l',k'} R^{(n,1)}(\z)^{l',k'}_{l_1,k_1} 
R^{(n,1)}(\zi)_{l',k'}^{l_2,k_2} = \d_{l_1,l_2}
 \d_{k_1,k_2},$
and crossing symmetry
$R^{(n,1)}(-q^{-1} \z)^{l',k'}_{l,k}=
R^{(n,1)}(\zi)^{n-l,k'}_{n-l',k}.$

\subsection{Elementary intertwiners}
First, we recall the definition of the elementary $U'$
intertwiners
\bac
\Phi(\z)&:&V(\L_i)\lra V(\L_{1-i})\ot V_{\z}^{(1)},\quad{\rm and}\\
\Psi^*(\z)&:&V_{\z}^{(1)}\ot V(\L_i) \lra V(\L_{1-i})\label{EVO}\ea
of \cite{JM,DFJMN}. Components are defined by
\bac
\Phi(\z) v&=&\sl_{k=0}^1\Phi_{k}(\z)v \ot u^{(1)}_{k},\label{TYPI}\\
\Psi^*(\z)(u^{(1)}_{k}\ot v)&=& \Psi^*_{k}(\z) v,\label{TYPII}\ea
where $v\in V(\L_i)$, and $\Phi_{k}(\z)$ and $\Psi^*_{k}(\z)$
are both maps $V(\L_i)\ra V(\L_{1-i})\ot \bC[\z,\zi]$. Note that $k=0,1$ here
corresponds to $\ep=+,-$ in \cite{JM}. 
These intertwiners are unique up to a normalisation, which is
fixed by the requirements
\[\begin{array}{lllll}\nn
\bra{\L_1}\Phi_1(\z)\ket{\L_0}=1,
\quad \bra{\L_0}\Phi_0(\z)\ket{\L_1}=1,\nn\\
\bra{\L_1}\Psi^*_0(\z)\ket{\L_0}=1,
\quad \bra{\L_0}\Psi^*_1(\z)\ket{\L_1}=1.
\nn\end{array}\]
Here we use the bra-ket notation, identifying $v_{\L_i}=\ket{\L_i}$.
\subsection{General intertwiners}
The existence and uniqueness of $U'$ intertwiners of the form
\be \Phi^{(n)}(\z) :V_{\z}^{(n)} \otimes V(\L_i) \lra V(\L_{1-i}) \otimes
V_{\z}^{(n+1)}\nn\ee
was demonstrated in \cite{Nak96}. Our construction differs from
that of \cite{Nak96} only in that we use the principal
evaluation representation. Following \cite{Nak96}, we construct
$\Phi^{(n)}(\z)$ in terms of the intertwiner $O^{(n)}
:V^{(1)}_{\xi_1}\ot
\cdots \ot V^{(1)}_{\xi_n}\ot V(\L_i)\lra 
V(\L_{1-i})\ot V^{(1)}_{\z_1}\ot \cdots \ot V^{(1)}_{\z_{n+1}}$ defined
by
\be O^{(n)} = {1 \ov f^{(n)}} 
\Phi(\z_1) \cdots
\Phi(\z_{n+1})\Psi^*(\xi_1)\cdots \Psi^*(\xi_n), \nn\ee
where
\bea f^{(n)}&=& \pl_{a}(-q^3 \xi_a^2)^{(n+1-a)/2}
\pl_{b}(-q^3 \z_b^2)^{(1-b)/2}\nn\\
&\times&
\pl_{a<a'} 
{\qp4{(\xi_{a'}/\xi_a)^2} \ov \qp4{q^2 (\xi_{a'}/\xi_a)^2}}
\pl_{b<b'} 
{\qp4{q^2 (\z_{b'}/\z_b)^2} \ov \qp4{q^4 (\z_{b'}/\z_b)^2}}
\pl_{a,b} 
{\qp4{q^3 (\xi_{a}/\z_b)^2} \ov \qp4{q(\xi_{a}/\z_b)^2}}.
\nn\ena
Here we adopt the convention that $a\in\{1,\cdots,n\}$ and
$b\in\{1,\cdots,n+1\}$.
We define $\Phi^{(n)}(\z)$ by \be
\Phi^{(n)}(\z) =
(1\ot \pi^{(n+1)}) O^{(n)}(\iota^{(n)}\ot 1)
\Big|_{\cal S},
\label{gint}\ee
where ${\cal S}$ is the one-dimensional submanifold in the parameter
space of $\{\xi_a,\z_b\}$:
\bea
{\cal S}&=&\{\xi_a=q^{-{n+1\ov 2} +a}\z;\ \z_b=q^{{n+2\ov 2}-b}\z \}.
\nn\ena
As we show in the appendix, each component of $O^{(n)}$ has poles
at ${\cal S}$. However, the combination
$(1\ot \pi^{(n+1)}) O^{(n)}(\iota^{(n)}\ot 1)$
is free of poles on ${\cal S}$.

Let $U'_i$ ($i=0,1$) be the $U'$ subalgebra generated by $e_i,f_i$
and $t_i$, and define 
\bea
\pi_i^{(n+1)}&:&V^{(1)}_{\zeta_1}\ot\cdots\ot V^{(1)}_{\zeta_{n+1}}
\rightarrow V^{(n+1)}_\zeta,\nn\\
\iota_i^{(n)}
&:&V^{(n)}_\zeta\rightarrow V^{(1)}_{\xi_1}\ot\cdots\ot V^{(1)}_{\xi_n}
\nn\ena
to be the unique $U'_i$ intertwiners normalised by
\bea
\pi^{(n+1)}_i(u^{(1)}_0\ot\cdots\ot u^{(1)}_0)&=&u^{(n+1)}_0,\nn\\
\iota^{(n)}_i(u^{(n)}_0)&=&u^{(1)}_0\ot\cdots\ot u^{(1)}_0.\nn
\ena

In order to prove that $\Phi^{(n)}(\z)$ defined by \mref{gint} is a 
$U'$ intertwiner,
it is enough to show that
\bea\label{RES1}
\Phi^{(n)}(\zeta)&=&(1\ot \pi_i^{(n+1)}) O^{(n)}(\iota_i^{(n)}\ot 1)
\Big|_{\cal S}.
\ena
The restriction to ${\cal S}$ in (\ref{RES1})
is regular, and so the rhs of (\ref{RES1}) is a $U'_i$ intertwiner.
In the appendix, we have shown that
if we divide the restriction to ${\cal S}$ into the two steps
\bea
\hbox{\it Step $1$}&:&\xi_a=q^\half\zeta_{n+2-a}\quad(1\leq a\leq
n),\quad {\rm and},\nn\\
\hbox{\it Step $2$}&:&\zeta_b=q^{{n+2\over2}-b}\zeta\quad(1\leq b\leq n+1),
\nn\ena
then each component of $O^{(n)}$ is regular.
The coefficients of the components in the linear combinations 
\mref{RES1} (for both
$i=0$ and $i=1$) and \mref{gint} coincide on ${\cal S}$.  
$\Phi^{(n)}(\z)$ is therefore an intertwiner for both $U'_0$ and
$U'_1$. This is a simple proof that $\Phi^{(n)}(\z)$ is a $U'$ 
intertwiner.

Define components by
\be 
\Phi^{(n)}(\z)(u_{l}^{(n)}\ot v)= \sl_{k=0}^{n+1} (\Phi^{(n)}_{l,k}(\z) v
\ot u_{k}^{(n+1)}).\label{gint2}\ee
The normalisation of \mref{gint} is such that
\be
\bra{\L_1} \Phi^{(n)}_{n,n+1}(\z) \ket{\L_0} =1,\quad
\bra{\L_0} \Phi^{(n)}_{0,0}(\z) \ket{\L_1} =1.\nn\ee

The intertwiner $\Phi^{(n)}(\z)$ has the following properties,
analogous to those
of the elementary intertwiner $\Phi(\z)$ given in \cite{JM} (indeed
we can identify $\Phi(\z)=\Phi^{(0)}(\z)$):
\bea 
g^{(n)} \sl_{k=0}^{n+1} \Phi^{(n)}_{l_1,k}(-q^{-1}\z )
\Phi^{(n)}_{n-l_2,n+1-k}(\z ) &=& \d_{l_1,l_2},\label{gunit}\\
\sl_{l',k'}R^{(1,n+1)}(\z/\xi)^{l',k'}_{l_1,k} 
\Phi_{l'}(\z) \Phi^{(n)}_{l_2,k'}(\xi) &=&
 \Phi^{(n)}_{l_2,k}(\xi) \Phi_{l_1}(\z),\label{gcommI}\\
\sl_{l',k'}\Psi^*_{k'}(\xi) \Phi^{(n)}_{l',k_1}(\z)
R^{(n,1)}(\z/\xi)^{l,k_2}_{l',k'} &=&
\Phi^{(n)}_{l,k_1}(\z) \Psi^*_{k_2}(\xi),\label{gcommII}\\
\xi^{-D} \Phi^{(n)}_{l,k}(\z) \xi^{D}&=& \Phi^{(n)}_{l,k}(\z/\xi)
.\label{gderiv}
\ena
where\vspace*{-2mm}
\be
g^{(n)}={\qp4{q^{2+2n}} \ov \qp4{q^{4+2n}}}.\nn\ee
In \mref{gderiv}, $D$ is the principal grading, which acts on
$V(\L_j)$  as
\be D (f_{i_1}f_{i_2} \cdots f_{i_N} v_{\L_j}) = 
N (f_{i_1}f_{i_2} \cdots f_{i_N} v_{\L_j}).\nn\ee
Relation \mref{gderiv} is a simple consequence of the analogous property
for $\Phi(\z)$ and $\Psi^*(\z)$ (see \cite{JM}).
Properties \mref{gunit}-\mref{gcommII} can be derived by
slightly modifying the proof of Theorem 5 in \cite{Nak96}.
We give a proof only of (\ref{gunit}).

We use the following intertwiner:
\bac
{\bar R}^{(n)}(\zeta_1,\cdots,\zeta_n)
:V^{(1)}_{\zeta_1}\ot\cdots\ot V^{(1)}_{\zeta_n}
&\rightarrow&V^{(1)}_{\zeta_n}\ot\cdots\ot V^{(1)}_{\zeta_1},\\[2mm]
{\bar R}^{(n)}(\zeta_1,\cdots,\zeta_n)(u^{(1)}_0\ot\cdots\ot u^{(1)}_0)
&=&u^{(1)}_0\ot\cdots\ot u^{(1)}_0,
\label{Rnot}\ea
and abbreviate ${\bar R}^{(n+1)}(\zeta_1,\cdots,\zeta_{n+1})$
to $R_\zeta$, and
${\bar R}^{(n)}(\xi_1,\cdots,\xi_n)$ to $R_\xi$.
We denote the restriction to
$\{\zeta_b=q^{{n+2\over2}-b}\zeta\}$ by $\Big|_\zeta$,
and
$\{\xi_a=q^{-{n+1\over2}+a}\zeta\}$ by $\Big|_\xi$.
We write the duality map as
\bea
C^{(n)}:V^{(n)}_{-q^{-1}\zeta}\ot V^{(n)}_\zeta&\rightarrow&\C,\nn\\
C^{(n)}(u^{(n)}_l\ot u^{(n)}_{n-l})&=&1.\nn
\ena
Then we have
\bea
(C^{(1)})^{n+1}(1\ot R_\zeta)\Big|_\zeta
&=&C^{(n+1)}(\pi^{(n+1)}\ot\pi^{(n+1)}),\label{L1}\\
(C^{(1)})^n(\iota^{(n)}\ot R_\xi\iota^{(n)})\Big|_\xi
&=&C^{(n)}.\label{L2}
\ena
The following properties of elementary intertwiners are also known \cite{JM}:
\bea
g^{(0)}\, C^{(1)} \Phi(-q^{-1}\z) \Phi(\z) &=& \id,
\label{unit}\\
\lim_{\xi_1 \ra -q^{-1} \xi_2} (1-q^{-2}\xi_2^2/\xi_1^2)\,
\Psi^*(\xi_1)\Psi^*(\xi_2)&=& g^{(0)} C^{(1)}.
\label{pole}
\ena

The proof of (\ref{gunit}) proceeds as follows:
In the notation of the appendix, the essential part of the lhs of (\ref{gunit})
is
\bea\label{PP}
\Bigl(\overline{\Phi(\zeta'_1)\cdots\Phi(\zeta'_{n+1})
\Psi^*(\xi'_1)\cdots\Psi^*(\xi'_n)}\Bigr)
\times\Bigl(\overline{\Phi(\zeta_1)\cdots\Phi(\zeta_{n+1})
\Psi^*(\xi_1)\cdots\Psi^*(\xi_n)}\Bigr).
\ena
To get (\ref{gunit}) we compose this with $C^{(n+1)}(\pi^{(n+1)}\ot\pi^{(n+1)})$
and $\iota^{(n)}\ot\iota^{(n)}$, and then restrict it to
${\cal S}$ and
$\{\xi'_a=q^{-{n+1\ov 2} +a}\z';\ \z'_b=q^{{n+2\ov 2}-b}\z'\}$,
and finally to $\zeta'=-q^{-1}\zeta$.
Let us denote this restriction by $\Big|_{\rm restrict}$.
Consider the product of operators
\bac\label{PP2}
&&\overline{\Phi(\zeta'_1)\cdots\Phi(\zeta'_{n+1})
\Psi^*(\xi'_1)\cdots\Psi^*(\xi'_n)\Phi(\zeta_1)\cdots\Phi(\zeta_{n+1})
\Psi^*(\xi_1)\cdots\Psi^*(\xi_n)}\\[2mm]
&=&\overline{\Phi(\zeta'_1)\cdots\Phi(\zeta'_{n+1})
\Phi(\zeta_1)\cdots\Phi(\zeta_{n+1})\Psi^*(\xi'_1)\cdots\Psi^*(\xi'_n)
\Psi^*(\xi_1)\cdots\Psi^*(\xi_n)}.
\ea
The two expressions (\ref{PP}) and (\ref{PP2}) are equal
up to a factor which is regular when we restrict in the way explained above.
Therefore, we can manipulate
\bea
C^{(n+1)}(\pi^{(n+1)}\ot\pi^{(n+1)})
\overline{\Phi(\zeta'_1)\cdots\Phi(\zeta'_{n+1})
\Phi(\zeta_1)\cdots\Phi(\zeta_{n+1})\Psi^*(\xi'_1)\cdots\Psi^*(\xi'_n)
\Psi^*(\xi_1)\cdots\Psi^*(\xi_n)}\nonumber\\
\times(\iota^{(n)}\ot\iota^{(n)})\Big|_{\rm restrict}\nn
\ena
instead of the expression containing (\ref{PP}).
Using (\ref{Rnot}) and (\ref{L1}), we reduce this expression to
\bea
(C^{(1)})^{n+1}
\overline{\Phi(\zeta'_1)\cdots\Phi(\zeta'_{n+1})
\Phi(\zeta_{n+1})\cdots\Phi(\zeta_1)\Psi^*(\xi'_1)\cdots\Psi^*(\xi'_n)
\Psi^*(\xi_n)\cdots\Psi^*(\xi_1)}
(\iota^{(n)}\ot R_\xi\iota^{(n)})\Big|_{\rm restrict}
\nn\ena
Removing the overline, using (\ref{unit}), (\ref{pole}) and \mref{L2},
and calculating the restriction of the contraction terms,
we arrive at (\ref{gunit}).

\setcounter{equation}{0}
\setcounter{section}{3}
\section{The Monodromy  Matrices}
In Section \ref{sechtm}, the results of perturbation theory and other considerations
led us to define the monodromy matrix as 
$T_{l,l'}^{(n+1)}(\z) = g^{(n)} \sl_{k=0}^{n+1} 
\Phi^{(n)}_{l,k} (\z) \ot \Phi^{(n)\
t}_{n-l',n+1-k}(\z)$. 
In Section \ref{secint}, following Nakayashiki, we defined
$\Phi^{(n)}(\z)$
and presented its properties \mref{gunit}-\mref{gderiv}. We now use
these results in order to derive certain properties of $T^{(n+1)}(\z)$.
\ss{Action on $\cF$}
One can view $\F=\cH\ot\cH^*$ as a linear map on $\cH$
via the canonical identification $\cH^*\ot\cH\simeq \End(\cH)$.
Then the action of $T_{l,l'}^{(n+1)}(\z)$ on $f\in \End(\cH)$
is given by
\be T_{l,l'}^{(n+1)}(\z) f = g^{(n)}\sl_{k=0}^{n+1} \Phi^{(n)}_{l,k} (\z)
\circ f \circ \Phi^{(n)}_{n-l',n+1-k}(\z) .\nn\ee

As an element of $\End(\cH)$, the vacuum in the $i$-th sector
was identified in \cite{JM} as 
\be \ket{vac}_{(i)} = \chi^{-\hf} (-q)^{D^{(i)}} P^{(i)},\nn\ee
where $P^{(i)}$ is the projector $\cH \ra V(\L_i)$, and 
$\chi=1/\qp4{q^2}$ is the principally specialised
character  of $V(\L_i)$. The superscript on the grading $D$ serves
only to indicate on which space $V(\L_i)$ it acts (we 
suppress the appearance of the projector from now on).

The action of $T_{l,l'}^{(n+1)}(\z)$ is given by
\bea
T_{l,l'}^{(n+1)}(\z) \ket{vac}_{(i)} &=& \chi^{-\hf}
g^{(n)}\sl_{k=0}^{n+1} \Phi^{(n)}_{l,k} (\z) (-q)^{D^{(i)}} 
\Phi^{(n)}_{n-l',n+1-k}(\z),\nn\\
&=& \chi^{-\hf}(-q)^{D^{(1-i)}}
g^{(n)}\sl_{k=0}^{n+1}  \Phi^{(n)}_{l,k} 
(-q^{-1}\z)\Phi^{(n)}_{n-l',n+1-k}(\z),\nn\\
&=& \d_{l,l'}\chi^{-\hf}(-q)^{D^{(1-i)}} =\d_{l,l'} \ket{vac}_{(1-i)}.
\nn\ena
Here we have used properties \mref{gderiv} and \mref{gunit}.

The Hamiltonian of the XXZ model is given by
\be H={(1-q^2) \ov 2q} \z{d\ov d\z} T^{(1)}(\z)\big|_{\z=1}.\nn\ee
Excited states are given by 
\be \ket {\xi_1,\cdots,\xi_m}_{\ep_1,\cdots,\ep_m;(i)}= 
(g^{(0)})^{-m/2} \chi^{-\hf}
\Psi^*_{\ep_1}(\xi_1) \cdots \Psi^*_{\ep_m}\, (\xi_1)
(-q)^{D^{(i)}}\nn\ee
with $|\xi_i|=1$ (see \cite{JM}). Using the commutation relation \mref{gcommII},
it is easy to show that the action of $T_{l,l'}^{(n+1)}(\z)$ 
on  the `m-particle state' 
$\ket {\xi_1,\cdots,\xi_m}_{\ep_1,\cdots,\ep_m;(i)}$ is given 
by
\bac
&&T_{l,l'}^{(n+1)}(\z) 
\ket {\xi_1,\cdots,\xi_m}_{\ep_1,\cdots,\ep_m;(i)}
= \\[1mm]
&&\sl_{\{\ep'_i,l_i\}} 
R^{(n,1)}(\z/\xi_1)^{l,\ep_1}_{l_1,\ep'_1}
R^{(n,1)}(\z/\xi_2)^{l_1,\ep_2}_{l_2,\ep'_2} \cdots
R^{(n,1)}(\z/\xi_m)^{l_{m-1},\ep_m}_{l',\ep'_m}
\ket {\xi_1,\cdots,\xi_m}_{\ep'_1,\cdots,\ep'_m;(1-i)}.\label{action}\ea
Here the sum is over $\ep'_1,\cdots,\ep'_m$ and $l_1,\cdots, l_{m-1}$.
If we represent the R-matrix graphically as in \cite{JM}, then this
action has the following rather simple representation:
\vspace*{5mm}

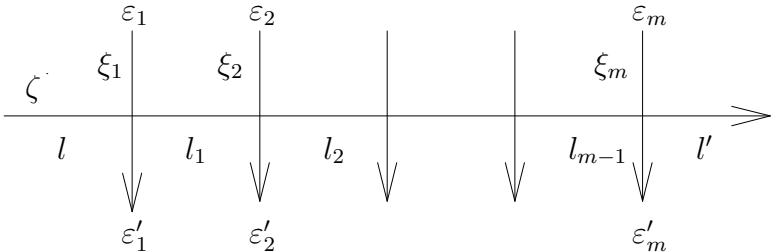
\setlength{\unitlength}{0.00083333in}
%
{\renewcommand{\dashlinestretch}{30}
\begin{picture}(5121,1535)(0,-10)
\drawline(287,1093)(287,1093)
\drawline(1615,1358)(1615,297)
\drawline(1555.000,537.000)(1615.000,297.000)(1675.000,537.000)
\drawline(3207,1358)(3207,297)
\drawline(3147.000,537.000)(3207.000,297.000)(3267.000,537.000)
\drawline(4003,1358)(4003,297)
\drawline(3943.000,537.000)(4003.000,297.000)(4063.000,537.000)
\drawline(22,828)(4800,828)
\drawline(4560.000,768.000)(4800.000,828.000)(4560.000,888.000)
\drawline(2411,1358)(2411,297)
\drawline(2351.000,537.000)(2411.000,297.000)(2471.000,537.000)
\drawline(819,1358)(819,231)
\drawline(759.000,471.000)(819.000,231.000)(879.000,471.000)
\put(150,950){\makebox(0,0)[lb]{\smash{{{$\z$ }}}}}
\put(1150,562){\makebox(0,0)[lb]{\smash{{{$l_1$ }}}}}
\put(2012,562){\makebox(0,0)[lb]{\smash{{{$l_2$}}}}}
\put(3539,562){\makebox(0,0)[lb]{\smash{{{$l_{m-1}$}}}}}
\put(287,562){\makebox(0,0)[lb]{\smash{{{ $l$}}}}}
\put(4335,562){\makebox(0,0)[lb]{\smash{{{$l'$}}}}}
\put(600,1093){\makebox(0,0)[lb]{\smash{{{$\xi_1$}}}}}
\put(1350,1093){\makebox(0,0)[lb]{\smash{{{$\xi_2$}}}}}
\put(3700,1093){\makebox(0,0)[lb]{\smash{{{$\xi_m$}}}}}
\put(752,1424){\makebox(0,0)[lb]{\smash{{{$\ep_1$}}}}}
\put(1548,1424){\makebox(0,0)[lb]{\smash{{{$\ep_2$}}}}}
\put(2411,1424){\makebox(0,0)[lb]{\smash{{{ }}}}}
\put(752,32){\makebox(0,0)[lb]{\smash{{{$\ep'_1$}}}}}
\put(1548,32){\makebox(0,0)[lb]{\smash{{{$\ep'_2$}}}}}
\put(3937,1424){\makebox(0,0)[lb]{\smash{{{$\ep_m$}}}}}
\put(3937,32){\makebox(0,0)[lb]{\smash{{{$\ep'_m$}}}}}
\end{picture}
}

\noindent This picture is related to the space of particles
and not to the co-ordinate lattice. For $n=0$, we have $T^{(1)}(\z)\ket
{\xi_1,\cdots,\xi_m}_{\ep_1,\cdots,\ep_m;(i)}=\tau(\z/\xi_1) \cdots 
\tau(\z/\xi_m)\ket
{\xi_1,\cdots,\xi_m}_{\ep_1,\cdots,\ep_m;(1-i)}$ as in \cite{JM},
where 
\be\tau(\z)=\zi {\qp4{q\z^2} \qp4{q^3 \z^{-2}}\ov \qp4{q\z^{-2}} 
\qp4{q^3 \z^{2}}}.\nn\ee

\ss{Commutation relations}

Making use of the explicit form of the action on $\cF$ given by
\mref{action}, and the Yang-Baxter equation \mref{YB}, 
we can immediately write down the commutation 
relationships of $T^{(n)}_{l,l'}(\z)$. 
\bea
\sl_{\bar{l}_1,\bar{l}_2}\bR^{(n,m)}(\z_1/\z_2)^{l_1,l_2}_{\bar{l}_1,\bar{l}_2} 
\,T^{(m+1)}_{\bar{l}_2,l'_2}(\z_2)
T^{(n+1)}_{\bar{l}_1,l'_1}(\z_1)&=&
\sl_{\bar{l}_1,\bar{l}_2}
T^{(n+1)}_{l_1,\bar{l}_1}(\z_1)
T^{(m+1)}_{l_2,\bar{l}_2}(\z_2)
\,\bR^{(n,m)}(\z_1/\z_2)^{\bar{l}_1,\bar{l}_2}_{l'_1,l'_2} 
, \nn\\
{[}T^{(n)}(\z_1),T^{(1)}(\z_2)]&=&0,\nn\ena
where $n,m\geq 1$. Thus $T^{(2)}_{l,l'}(\z)$ can be interpreted as the
$L$ operator of the spin-1/2 XXZ model in the infinite volume limit.

\ss{Fusion}
If we rewrite each of the $R^{(n,1)}(\z)$
that occur on the
rhs of \mref{action} using the fusion expression \mref{fusion},
then we obtain a fusion
relation for $T^{(n+1)}(\z)$ in terms of $T^{(2)}(\z)$ and 
$T^{(1)}(\z)$. 

Consider the operator $F$ which counts the number of particles
in the particle picture.
Define 
\bea
{\bar T}^{(n+1)}(\zeta)
&=&\cases{T^{(n+1)}(\zeta)T^{(1)}(\zeta)^{-1}&if $n\equiv0\bmod4$;\cr
T^{(n+1)}(\zeta)&if $n\equiv1\bmod4$;\cr
T^{(n+1)}(\zeta)T^{(1)}(\zeta)&if $n\equiv2\bmod4$;\cr
(-1)^{F}T^{(n+1)}(\zeta)&if $n\equiv3\bmod4$.\cr}
\label{corr}\ena
We note that there is an equality
$
{\bar T}^{(n+1)}(\zeta)=
T^{(n+1)}(\zeta)\pl_{a=0}^{n-2}T^{(1)}(q^{{n-2\over2}-a}\zeta).
$
Then, we have the following fusion relation:
\bea
{\bar T}^{(n+1)}_{l,l'}(\z) &=& {\g^{(n)}_{l'}  \ov \g^{(n)}_{l} }
\sl_{l'_1+\cdots+l'_n=l'} 
T^{(2)}_{l_1,l'_1}(\z q^{n-1 \ov 2})
T^{(2)}_{l_2,l'_2}(\z q^{n-3 \ov 2})
\cdots T^{(2)}_{l_n,l'_n}(\z q^{1-n \ov 2}),\nn\ena
where $n\geq1$. Here, the $l_a$ are specified only by the requirement
 $l_1+\cdots+l_n=l$;
the formula is independent of the actual choice of $l_a$.
\newpage
\section{Discussion}
In this paper, we have studied the $U_q(\widehat{sl}_2)$ symmetry of
the spin $\half$ XXZ model in the massive regime by making use of
Nakayashiki's intertwiners. We have conjectured that in the infinite volume
limit the half transfer matrix with spin $\half$ quantum space and
spin ${n+1\over 2}$ auxiliary space is represented by the intertwiner
\bea\label{NVO}
\Phi^{(n)}(\zeta)&:&V^{(n)}_\zeta\otimes V(\Lambda_i)
\rightarrow V(\Lambda_{1-i}\otimes V^{(n+1)}_\zeta).
\ena
This implies, in particular, that the monodromy matrix with spin $1$ 
auxiliary space enjoys the commutation relations of the $L$ operator.

In \cite{Nak96}, Nakayashiki uses the operator $\Phi^{(n)}(\zeta)$
to diagonalise the spin $\half$ XXZ model with higher spin impurities.
In the language of the six vertex model, this is equivalent to inserting
lines with higher spin. The difference between our approach and Nakayashiki's
is that we consider the monodromy matrices which are parallel to the inserted
lines, while Nakayashiki considers the transfer matrix which is perpendicular
to them. In Nakayashiki's case the space $V^{(n)}_\zeta$ in (\ref{NVO})
corresponds to the degeneracy of the vacuum states of the transfer matrix.
In our case, we have found that the same space corresponds to the boundary
conditions for the monodromy matrices.

We have derived the fusion relation for the monodromy matrices.
It almost corresponds to the fusion construction of the space $V^{(n)}_\zeta$
in (\ref{NVO}) out of the spaces with $n=1$. However, we have found that the
monodromy matrices contain the correction factor given in
\mref{corr}, which is diagonal in each
irreducible $m$ particle representation in the physical space of states.
\subsubsection*{Acknowledgements}
{\small TM thanks everyone in Durham, in particular,
Ed Corrigan and his family, for their hospitality
during the period in which this work was begun.
He also thanks Masaki Kashiwara for useful discussions.
RW thanks everyone in RIMS, and 
acknowledges the RIMS/JSPS/Isaac Newton Inst./Royal Society
exchange scheme for supporting his stay in Kyoto. He is also pleased
to acknowledge the EPSRC for providing him with an advanced fellowship,
and TMR network \#ERBFM-RXCT960012 for funding his stay in Paris.
Last but not least, the authors would like to thank the Centre Emile Borel
and the organisers of the Semestre Sytemes Integrable for their hospitality.}
\pagebreak

\newpage
\appendix
\setcounter{equation}{0}
\setcounter{section}{0}
\section{Regularity of Matrix Elements}
In this appendix, we prove a few statements concerning the regularity of the
matrix elements of the product of type I and type II
vertex operators.
We start from the bosonization of vertex operators given on page
140 of \cite{JM}.
Consider the product of vertex operators
\bea
O&=&\Phi_{k_1}(\z_1)\cdots\Phi_{k_m}(\z_m)
\Psi^*_{l_1}(\xi_1)\cdots\Psi^*_{l_n}(\xi_n).
\nn\ena
It contains integrals with respect to the variables
$w_{\bar b}$ in $X^-(w_{\bar b})$, for ${\bar b}$ such that $k_{\bar b}=0$,
and $u_{\bar a}$ in $X^+(u_{\bar a})$, for ${\bar a}$ such that $l_{\bar a}=0$.

After normal-ordering, the integrand,
which depends on the variables $\zeta_b$, $\xi_a$,
$w_{\bar b}$ and $u_{\bar a}$, consists of the following three parts:

(i) $O_1$, the contraction terms,
pairwise in $\Phi_1\Phi_1$, $\Phi_1\Psi^*_1$, $\Psi^*_1\Phi_1$ or $\Psi^*_1\Psi^*_1$.

We have
\bea
O_1&=&
\prod_{b<b'}(-q^3\zeta_b^2)^\half
{\infp{q^2\zeta_{b'}^2/\zeta_b^2}\over\infp{q^4\zeta_{b'}^2/\zeta_b^2}}
\prod_{a<a'}(-q^3\xi_b^2)^\half
{\infp{\xi_{a'}^2/\xi_a^2}\over\infp{q^2\xi_{a'}^2/\xi_a^2}}
\nonumber\\&\times&
\prod_{a,b}(-q^3\zeta_b^2)^{-\half}
{\infp{q^3\xi_a^2/\zeta_b^2}\over\infp{q\xi_a^2/\zeta_b^2}}
.\nn\ena

This is a function of $\zeta_b$, $\xi_a$.
In the above setting, pairs of the form $\Psi^*_1\Phi_1$
do not appear because $\Phi_1$ is always in the left of $\Psi^*_1$;

(ii) $O_2$, the contraction terms for the rest of the pairs.

We have
\bea
&&\prod_{{\bar b}<{\bar b}'}
(w_{\bar b}-w_{{\bar b}'})(w_{\bar b}-q^2w_{{\bar b}'})
\prod_{{\bar a}<{\bar a}'}
(u_{\bar a}-u_{{\bar a}'})(u_{\bar a}-q^{-2}u_{{\bar a}'})
\nonumber\\&\times&
\prod_{{\bar a},{\bar b}}
{1\over(w_{\bar b}-qu_{\bar a})(w_{\bar b}-q^{-1}u_{\bar a})}
\prod_{{\bar a},b}(u_{\bar a}-q^3\zeta_b)
\prod_{a,{\bar b}}(w_{\bar b}-q^3\xi_a)\nonumber\\&\times&
\prod_{a\leq{\bar a}}{q\over u_{\bar a}-q^4\xi_a^2}
\prod_{{\bar a}\leq a}{1\over u_{\bar a}-q^2\xi_a^2}
\prod_{b\leq{\bar b}}{1\over q(w_{\bar b}-q^2\zeta_b^2)}
\prod_{{\bar b}\leq b}{1\over w_{\bar b}-q^4\zeta_b^2}
;\nn\ena

(iii) $O_3$, the rest, which is a normal-ordered product
of vertex operators with coefficients that are Laurent polynomials
in $\zeta_b$, $\xi_a$, $w_{\bar b}$ and $u_{\bar a}$.

The contours for the integrals are such that
the $q^4\zeta^2_b$, $qu_{\bar a}$, $q^{-1}u_{\bar a}$
are inside, and the $q^2\zeta^2_b$ are outside
of the contour for the $w_{\bar b}$ integration;
the $q^2\xi^2_a$ are
inside, and the $q^4\xi^2_a$, $qw_{\bar b}$, $q^{-1}w_{\bar b}$ are outside
of the contour for the $u_{\bar a}$ integration.

Denote the quantity which is $O$ with $O_1$ removed, by $\overline O$.
Note that because of the commutation relation (A.3) of \cite{JM},
$\Phi_{l_b}(\zeta_b)$ commutes with $\Psi^*_{k_a}(\xi_a)$
inside of $\overline O$.

Now let us examine the regularity of the matrix elements of $\overline O$.
It is enough to consider the $O_2$ term in the integrand.
The possible pinchings of the contours occur in the following four
cases (which we list  with the relevant factors in the integrand):
\bea
&\hbox{\it Case $1$}&
\quad{1\over(u_{\bar a}-q^4\xi_{a_1}^2)(u_{\bar a}-q^2\xi_{a_2}^2)}
\;\hbox{\rm \, at \,$\xi_{a_1}^2=q^{-2}\xi_{a_2}^2$, \,for\, $a_1\leq{\bar a}\leq a_2$,}
\nn\\
&\hbox{\it Case $2$}&
\quad{1\over(w_{\bar b}-q^2\zeta_{b_1}^2)(w_{\bar b}-q^4\zeta_{b_2}^2)}
\;\hbox{\rm \, at \,$\zeta_{b_1}^2=q^2\zeta_{b_2}^2$, \,for\, $b_1\leq{\bar b}\leq b_2$,}
\nn\\
&\hbox{\it Case $3$}&
\quad{w_{\bar b}-q^3\xi_a^2\over(w_{\bar b}-q^2\zeta_b^2)
(w_{\bar b }-qu_{\bar a})(u_{\bar a}-q^2\xi_a^2)}
\;\hbox{\rm \, at\, $\zeta_{b_1}^2=q\xi_a^2$, \,for\, $b\leq{\bar b}$, ${\bar a}\leq a$,}
\nn\\
&\hbox{\it Case $4$}&
\quad{u_{\bar a}-q^3\zeta_b^2\over(w_{\bar b}-q^2\zeta_b^2)
(w_{\bar b }-q^{-1}u_{\bar a})(u_{\bar a}-q^2\xi_a^2)}
\;\hbox{\rm  \,at \,$\zeta_{b_1}^2=q^{-1}\xi_a^2$,
\,for\, $b\leq{\bar b}$, ${\bar a}\leq a$.}\nonumber
\ena

Cases 1 and 2 give rise to poles. 
They are at most simple because of the factor
\newline $\prod_{{\bar a}<{\bar a}'}(u_{\bar a}-u_{{\bar a}'})$ or
$\prod_{{\bar b}<{\bar b}'}(w_{\bar b}-w_{{\bar b}'})$, respectively.
Cases 3 and 4 are pole free because of the factor
$w_{\bar b}-q^3\xi_a^2$ or $u_{\bar a}-q^3\zeta_b^2$, respectively.
The final remark is that the restriction of $\overline O$
at $q\zeta_b^2=\xi_{a_2}^2$ for any $b$ is regular
at $\xi_{a_1}^2=q^{-2}\xi_{a_2}^2$, and the restriction
at $q\zeta_{b_2}^2=\xi_a^2$ for any $a$ is regular
at $\zeta_{b_1}^2=q^2\zeta_{b_2}^2$.
This is because of the following factors in the numerator:
\bea
u_{\bar a}-q^3\zeta_b^2
&=&(u_{\bar a}-q^2\xi_{a_2}^2)
+q^2(\xi_{a_2}^2-q\zeta_b^2),\nn\\
w_{\bar b}-q^3\xi_a^2
&=&(w_{\bar b}-q^4\zeta_{b_2}^2)
+q^3(q\zeta_{b_2}^2-\xi_a^2).\nn
\ena

\begin{thebibliography}{1}

\bibitem{FM}
O.~Foda and T.~Miwa.
\newblock {Corner Transfer Matrices and Quantum Affine Algebras}.
\newblock {\em Int. J. Mod. Phys.}, A7 supplement 1A:279, 1992.

\bibitem{DFJMN}
B.~Davies, O.~Foda, M.~Jimbo, T.~Miwa, and A.~Nakayashiki.
\newblock {Diagonalisation of the XXZ Hamiltonian by vertex operators}.
\newblock {\em Comm. Math. Phys.}, 151:89--153, 1993.

\bibitem{JMMN}
M.~Jimbo, K.~Miki, T.~Miwa, and A.~Nakayashiki.
\newblock {Correlation Functions of the {X}{X}{Z} model for {${\D}<-1$}}.
\newblock {\em Phys. Lett.}, A168:256--263, 1992.

\bibitem{JMN}
M.~Jimbo, T.~Miwa, and A.~Nakayashiki.
\newblock Difference equations for the correlation functions of the
  eight-vertex model.
\newblock {\em J. Phys. A}, 26:2199--2209, 1993.

\bibitem{JM}
M.~Jimbo and T.~Miwa.
\newblock {\em Algebraic Analysis of Solvable Lattice Models}.
\newblock CBMS Regional Conference Series in Mathematics vol. 85, AMS, 1994.

\bibitem{Miki}
K.~Miki.
\newblock {Creation/annihilation operators and form factors of the XXX model}.
\newblock {\em Phys. Lett.}, 186:217--224, 1994.

\bibitem{Nak96}
A.~Nakayashiki.
\newblock {Fusion of q-Vertex Operators and its Application to Solvable Vertex
  Models}.
\newblock {\em Comm. Math. Phys.}, 177:27--62, 1996.

\end{thebibliography}
\end{document}